\documentclass{ieeeaccess}
\usepackage{cite}
\usepackage{array}
\usepackage{float}
\usepackage{multirow}
\usepackage{hyperref}
\usepackage{amsmath,amssymb,amsfonts}
\usepackage[utf8]{inputenc}
\usepackage{amsmath,amssymb,amsfonts}
\usepackage{algorithmic}
\usepackage{graphicx}
\usepackage{textcomp}

\usepackage{bm}
\makeatletter
\AtBeginDocument{\DeclareMathVersion{bold}
\SetSymbolFont{operators}{bold}{T1}{times}{b}{n}
\SetSymbolFont{NewLetters}{bold}{T1}{times}{b}{it}
\SetMathAlphabet{\mathrm}{bold}{T1}{times}{b}{n}
\SetMathAlphabet{\mathit}{bold}{T1}{times}{b}{it}
\SetMathAlphabet{\mathbf}{bold}{T1}{times}{b}{n}
\SetMathAlphabet{\mathtt}{bold}{OT1}{pcr}{b}{n}
\SetSymbolFont{symbols}{bold}{OMS}{cmsy}{b}{n}
\renewcommand\boldmath{\@nomath\boldmath\mathversion{bold}}}
\makeatother

\def\BibTeX{{\rm B\kern-.05em{\sc i\kern-.025em b}\kern-.08em
    T\kern-.1667em\lower.7ex\hbox{E}\kern-.125emX}}

\begin{document}

\title{Automated Quantum Software and AI Engineering}
\author{\uppercase{Nazanin Siavash}\authorrefmark{1} and
\uppercase{Armin Moin}\authorrefmark{2}}

\address[1]{Department of Computer Science, University of Colorado Colorado Springs (UCCS), Colorado Springs, Colorado, USA (e-mail: nsiavash@uccs.edu)}
\address[2]{Department of Computer Science, University of Colorado Colorado Springs (UCCS), Colorado Springs, Colorado, USA (e-mail: amoin@uccs.edu))}

\markboth{Siavash and Moin: Automated Quantum Software and AI Engineering}
{Siavash and Moin: Automated
 Quantum Software and AI Engineering}

\corresp{Corresponding author: Armin Moin (e-mail: amoin@uccs.edu)}

\begin{abstract}
In this paper, we conduct a systematic literature review of (semi-) automated approaches to Quantum Software Engineering (QSE) and Quantum Artificial Intelligence (QAI). Prior work in the literature indicated that both Software Engineering (SE) and Artificial Intelligence (AI) practices may become more efficient by using (semi-) automated approaches. This also holds in the Quantum Computing (QC), Quantum Information Science (QIS), and Quantum Engineering (QE) world, as well as in hybrid quantum-classical applications. In fact, automation is even more crucial in such cases since there is a limited number of developers and AI experts (e.g., data scientists) who possess the required knowledge and skills in QC. Moreover, in hybrid setups, automation may help decide what part of the application should be deployed on quantum hardware and on which of the available quantum platforms, if applicable. This can be a significant help to achieve productivity leap and efficiency even for subject matter experts. Unlike prior literature reviews and surveys, this work focuses on automation in SE and AI for quantum and hybrid quantum-classical applications and identifies the recent trends and future directions through a systematic literature review. We are interested in methods and techniques that can enable a broader development and deployment of quantum and hybrid AI-enabled software systems.
\end{abstract}

\begin{keywords}
quantum computing, artificial intelligence, software engineering, automation
\end{keywords}

\titlepgskip=-21pt

\maketitle

\section{Introduction}
\label{sec:introduction}
\PARstart{Q}{uantum} Computing (QC) is rapidly evolving from a theoretical supremacy into a practical advantage for various applications, presenting opportunities to overcome some limitations of classical computation. For lots of applications, classical computation will still be appropriate and meaningful for the foreseeable future. However, in specific cases, the advantage that can be achieved by deploying QC, Quantum Information Science (QIS), and Quantum Engineering (QE) will be disproportionately significant. There are not many algorithms that are \textit{quantum} by design, known as \textit{quantum algorithms}, meaning that they can benefit from the particular characteristics of quantum computers. Additionally, quantum data, which means data stored in qubits rather than bits, is yet another dimension of the discussion. One may use quantum algorithms with or without quantum data. Automation can touch on various aspects of the Software Engineering (SE) and Artificial Intelligence (AI) processes and practices. Moreover, automation may be partial or full and can occur at different levels. For instance, having a piece of technology that can automate or assist in choosing the appropriate mode, namely quantum, classical, or hybrid quantum-classical computation, as well as the other options, such as the model of computation (e.g., gate-based vs. adiabatic), for a specific computational task or application would be interesting and beneficial to the SE and AI communities.

As more software systems are becoming AI-enabled (e.g., machine learning-enabled), we are witnessing more overlap and synergies between the SE and AI communities. Similarly, as the QC, QIS, and QE fields are advancing, there will be more need for studying Quantum Software Engineering (QSE) and Quantum AI (QAI) in a holistic manner. Therefore, this study brings all the mentioned fields together and conducts the first Systematic Literature Review (SLR) of the emerging research and development areas of QSE and QAI, as well as hybrid quantum-classical SE and AI, with a focus on automation in the SE and AI processes and practices. We do not differentiate between the works that have enabled full or partial (i.e., semi) automation.

Hence, the contribution of this work is providing a comprehensive SLR to the scientific community featuring automation in QSE (e.g., automated code generation or automated testing) and QAI (e.g., Automated Machine Learning, known as AutoML). While there exist prior works in the literature that surveyed or reviewed the literature in QSE \cite{Dwivedi+2024}, QAI \cite{YuZhao2023, Melnikov+2023, Klusch2024}, or Automated QSE (AQSE) \cite{Kusyk+2021, Sarkar2024}, this paper presents the first SLR that encompasses all these topics, as well as Automated QAI (AQAI) in a single study in a holistic manner. We begin by introducing a number of fundamental concepts and continue with the specific areas, namely QSE, QAI, AQSE, and AQAI, highlighting relevant studies in each domain according to the research methodology. Although AQSE and AQAI are still new, bringing together current efforts at this early stage can help guide future research, make terms clearer, and avoid repeating the same work. Additionally, we identify recent trends, discuss key challenges, and outline future research directions for each of the recent relevant works individually.

The remainder of this paper is structured as follows: Section \ref{sec:background} provides some background information. In Section \ref{sec:related-work}, related work, that is, (systematic) literature reviews and survey papers in the target fields, are explored. Additionally, our research methodology is explained in Section \ref{sec:methodology}. Sections \ref{sec:qse} to \ref{sec:aqai} discuss the reviewed works in the literature in the areas of QSE, QAI, AQSE, and AQAI, respectively, with detailed comparative analysis, as well as the identified recent trends, challenges, and future directions individually. Finally, Section \ref{sec:conclusion-future-directions} concludes and outlines overall future research directions.
\section{Background}\label{sec:background}
In this section, we review a number of fundamental concepts, namely, Automated SE, Automated AI, QC/QIS/QE, as well as Quantum SE and Quantum AI.

\subsection{Automated SE}
Automated SE in the classical (i.e., non-quantum) world has various facets. It can range from automated generation of a skeleton of the program to automated verification (e.g., testing or even formal verification) and validation, or beyond that to automated generation of a complete solution \cite{Moin+2022-ICSE}. The ultimate goal is to enhance the efficiency of software developers and other practitioners, such as testers, through abstraction and automation, as well as improve the quality of the produced software system through automation. Various methods and techniques, such as deploying Large Language Models (LLMs), can enable automation in SE. Last but not least, there has been a growing interest in automating SE for AI-enabled systems \cite{Raedler+2024}. Note that regardless of whether AI is used in enabling the automation process in SE or not, the target software system may or may not be an AI-enabled system.

\subsection{Automated AI}
AI has various sub-disciplines, including but not limited to machine learning, logic programming, and multi-agent systems. Automation in AI can help practitioners working in AI fields, for example, data scientists and machine learning engineers working on Data Analytics and machine learning projects, become more efficient and productive in their tasks. In the case of machine learning, this automated approach is called AutoML. The automated fine-tuning of hyperparameters of the machine learning model or the construction and evaluation of several model architectures and configurations with the aim of choosing the best-performing model in an automated and efficient manner can often enable practitioners to achieve breakthroughs in their work (e.g., see \cite{Moin+2025-ITNG}).

\subsection{Quantum Computing, Quantum Information Science, and Quantum Engineering}
\begin{figure}[h]
\centering
\includegraphics[width=0.65\columnwidth]{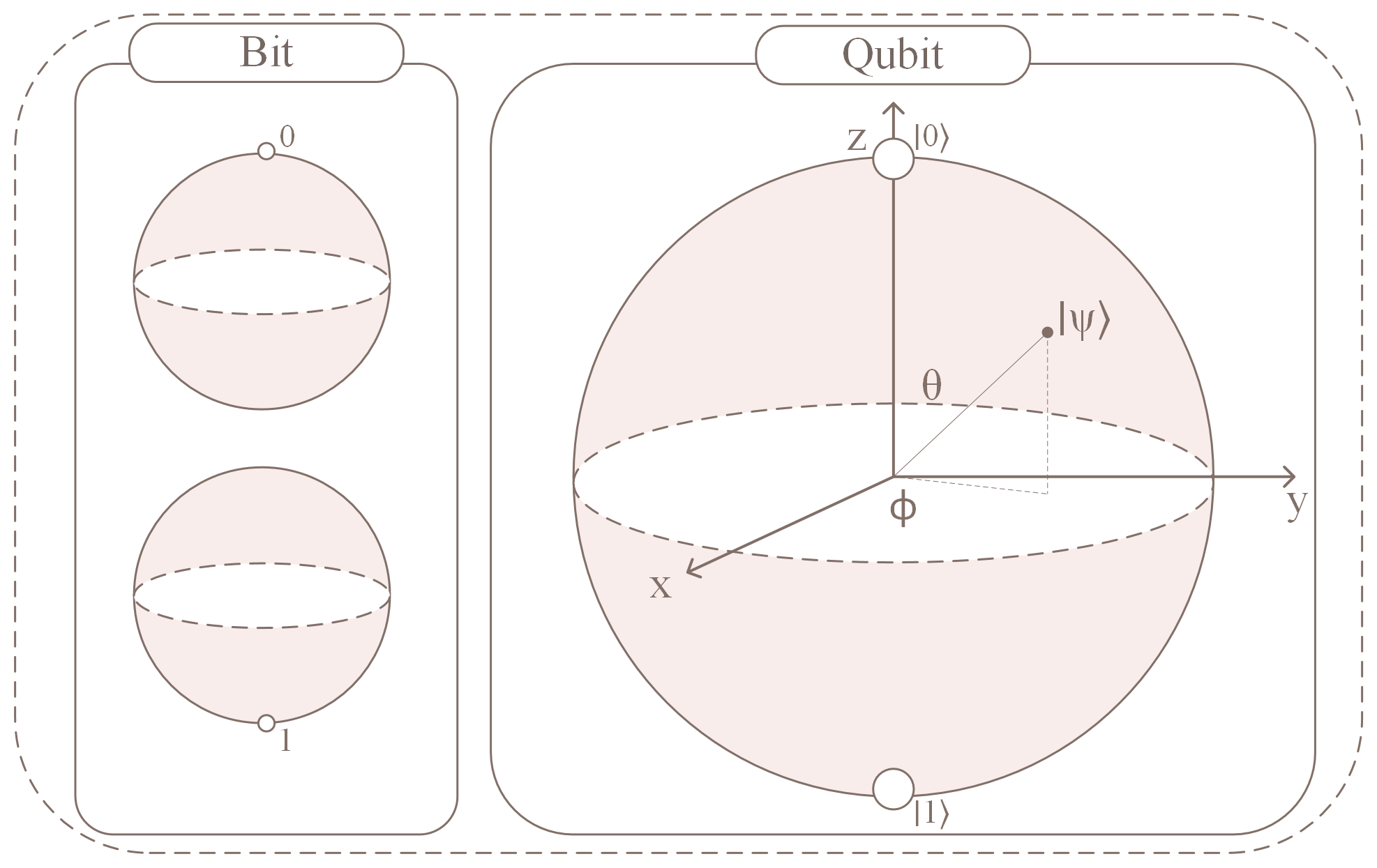}\caption{Illustration of a bit with a determinate state of zero or one vs. a qubit (quantum bit) in a superposition state.}
\label{fig:bit-qubit}
\end{figure}

Various fields, including AI and SE, are undergoing rapid transformations driven by disruptive quantum technologies. These can offer new possibilities for solving some of the complex problems deemed challenging by classical (i.e., non-quantum) computers in a reasonable amount of time. QC involves executing computations based on the principles of quantum mechanics \cite{Deutsch1985,Dirac1981,GyongyosiImre2019}, a field historically explored by physicists rather than computer scientists or software engineers \cite{Zhao2020}. Since the early 20th century, quantum theory has led to the development of quantum-based technologies, impacting various fields such as cryptography, superconductors, and computation \cite{Fernandez+2022}. The Applied Physics side of QC focuses on the quantum properties of specific subatomic particles, such as photons and electrons, which are utilized for calculations and extensive information processing \cite{Peral+2024}. The theoretical basis for QC dates back to the 1980s \cite{Feynman2002} when the integration of quantum mechanics and information theory led to the emergence of quantum information theory. The advancement of QC was driven by the belief that specific quantum phenomena could enhance computing speeds beyond the limits of classical systems. These benefits stem from leveraging quantum characteristics \cite{Bawa2024}, which can enable quantum computers to outperform classical computers in both processing time and cost efficiency. In QC, the fundamental unit of data is known as a quantum bit, called qubit \cite{Shafique+2024}. One key characteristic of QC is \textit{superposition}; when a qubit is in a superposition \cite{BennettShor1998}, it occupies multiple states at once. Thus, its state remains indeterminate until a measurement is made. The illustration of a qubit in a superposition state and a classical bit in a determinate state of zero or one is shown in Figure \ref{fig:bit-qubit}. Additionally, the mathematical representation of this superposition is provided in Equation \ref{eq:superposition1}.

\begin{equation}
\label{eq:superposition1}
|\psi\rangle=\alpha|0\rangle+\beta|1\rangle
\end{equation}
where $|\psi\rangle$ is the quantum state, and $\alpha$ and $\beta$ are probability amplitudes.

Considering \( |0\rangle \) as \( \begin{bmatrix} 0 \\ 0 \end{bmatrix} \) and \( |1\rangle \) as \( \begin{bmatrix} 0 \\ 1 \end{bmatrix} \), Equation \ref{eq:superposition1} can be written as:

\begin{equation}
\label{eq:superposition2}
|\psi\rangle =  \begin{bmatrix}
 \alpha \\
0 
\end{bmatrix} + \begin{bmatrix}
0 \\
\beta 
\end{bmatrix} = \begin{bmatrix}
\alpha \\
\beta 
\end{bmatrix}
\end{equation}

The square of the modulus of each amplitude indicates the likelihood of observing the corresponding state, and the total of these squared moduli across all states must equal one, as can be seen in Equation \ref{eq:superposition3}. 
\begin{equation}
\label{eq:superposition3}
|\alpha|^2 + |\beta|^2 = 1
\end{equation}

Besides superposition, another important QC characteristic is \textit{entanglement}. Pairs of qubits can become \textit{entangled}. This means that both entangled qubits share a single quantum state, and altering the state of one qubit will instantaneously affect the state of the other in a predictable manner, regardless of the distance separating them \cite{Sutor2019}. 

Well-known algorithms such as Grover's \cite{Grover1996}, Shor's \cite{Shor1994}, and Simon's \cite{Simon1997} leverage these quantum characteristics to accomplish some tasks that are challenging for classical computation to achieve.
There exist different \textit{models of computation} in QC. \textit{Gate-based} (also known as \textit{circuit-based}) and \textit{adiabatic} (closely related to \textit{quantum annealing} and often used interchangeably) are two popular ones \cite{Massoli+2022}. Gate-based QC operates based on quantum gates, which are analogues to digital logic gates in classical computing. Quantum gates manipulate quantum information similar to classical gates that can manipulate classical inputs according to their specification. NOT, Hadamard, T-gate, Conjugate transpose of T-gate, U3-gate, Toffoli, and CNOT are some common quantum gates \cite{Kusyk+2021}. In contrast, adiabatic QC \cite{Albash+2018} relies on the principles of the adiabatic theorem \cite{Kato1950}. In this model of computation, the quantum system representing the search space of a computational problem slowly transitions from a simple initial state to a final state that encapsulates the solution. The decision to use one quantum model of computation over another should typically be influenced by the specific problem being tackled and the available hardware resources.

\begin{figure}[h]
\centering
\includegraphics[width=0.45\columnwidth]{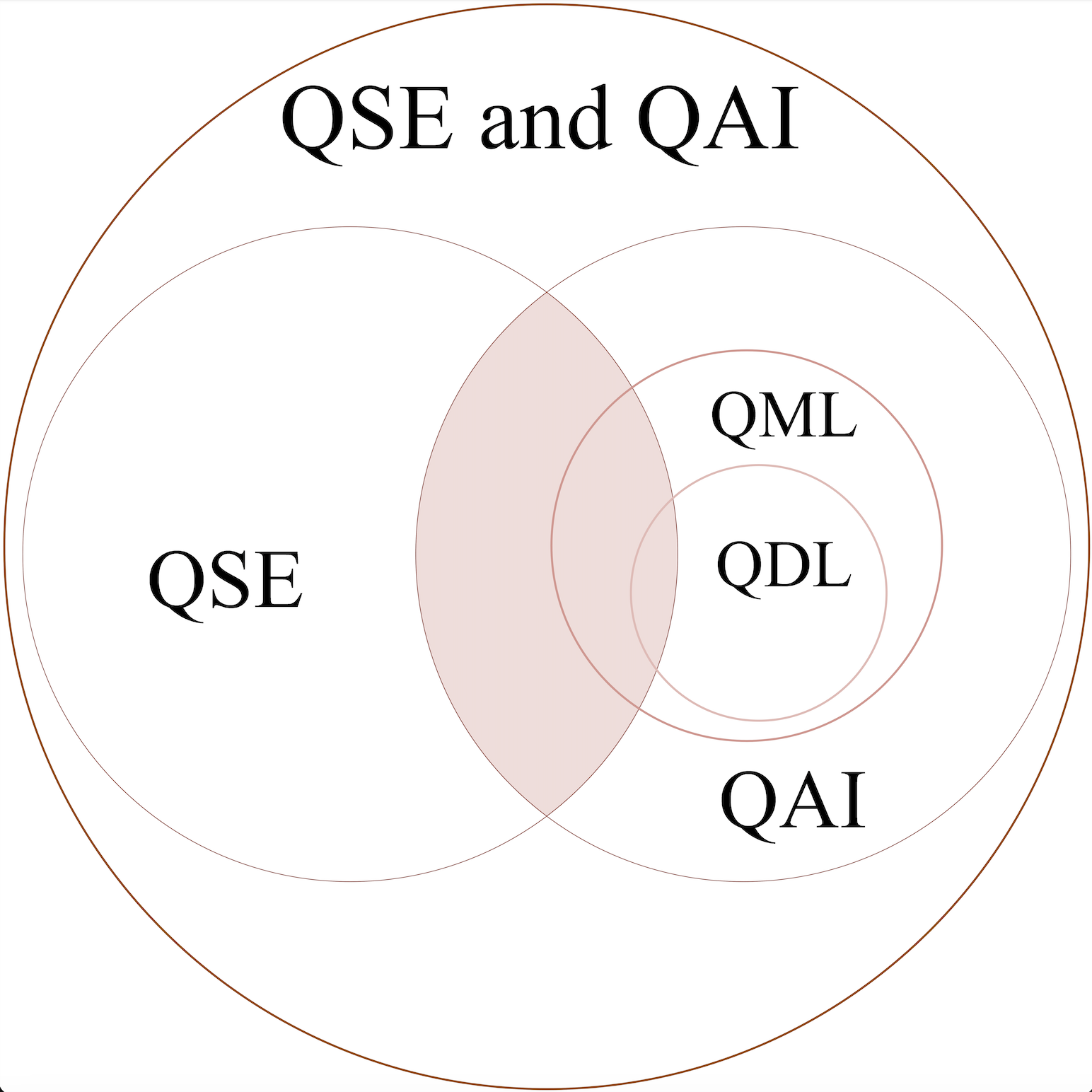}\caption{A Venn diagram illustrating the overlap of QSE and QAI to enable smart quantum-classical software systems.}
\label{fig:topics-quantum-se-ai}
\end{figure}

Finally, similar to Computer Science, Information Technology/Science, and Computer Engineering, which are centered around computing but have slightly different focuses and various theoretical or applied levels, QC, QIS, and QE are all centered around computing on quantum processors but with a slightly different concentration. As mentioned above, QC has an applied Physics facet. However, it also includes a Computer Science aspect. QIS is relevant to the latter. Additionally, QE brings QIS and engineering disciplines together. For instance, designing and developing quantum sensor technologies lies within the scope of QE \cite{MIT-QE}. 

\subsection{Quantum Software Engineering and Quantum Artificial Intelligence}
SE has evolved through cycles of innovation, from early computing foundations in the 1930s to structured programming, object-oriented methodologies, and modern DevOps practices \cite{Serrano+2022, Booch2018}. Each era has brought advancements, from high-level languages to service-oriented computing \cite{PapazoglouGeorgakopoulos2003}, driven by hardware advancement and algorithm progress, as well as emerging needs. Today, QC is poised to lead us into a disruptive age of SE, necessitating both adaptations of some of the classical software systems and the development of novel quantum-classical ones. However, effective operationalization of QC requires robust SE methodologies, processes, and tools \cite{Akbar+2024}. This emerging area of SE is called \textit{Quantum SE (QSE)}. As quantum technology advances, the demand for quantum-enabled software-intensive approaches tailored to industrial applications will grow. QSE is emerging as a critical discipline, requiring software engineers to adapt and collaborate with quantum physicists and researchers to develop hybrid classical-quantum solutions that bridge the gap between theory and real-world implementation \cite{Dwivedi+2024}. However, with the recent advances in high-level frameworks and libraries, such as the IBM Qiskit \cite{IBM-Qiskit}, software developers can become more self-sufficient if they target the respective supported hardware. Furthermore, prior studies emphasized that quantum software and hardware should ideally be co-designed. Certain algorithms and specific hardware technologies and models of computation perform significantly better for particular tasks \cite{Moin+2022-MDE4QAI}. Last but not least, the synergy between AI and QC opens up a novel frontier in intelligent systems, reshaping our understanding of computing \cite{PrezPrieta2022}. QC has fundamentally altered our perspective on computation, prompting a reassessment of many areas within computer science, including AI.

Figure \ref{fig:topics-quantum-se-ai} depicts the overlap between QSE and QAI. Note that automated approaches enabling solutions at the intersection of QSE and QAI are the main focus of this study.

\begin{figure*}[h]
\centering
\includegraphics[width=1.8\columnwidth]{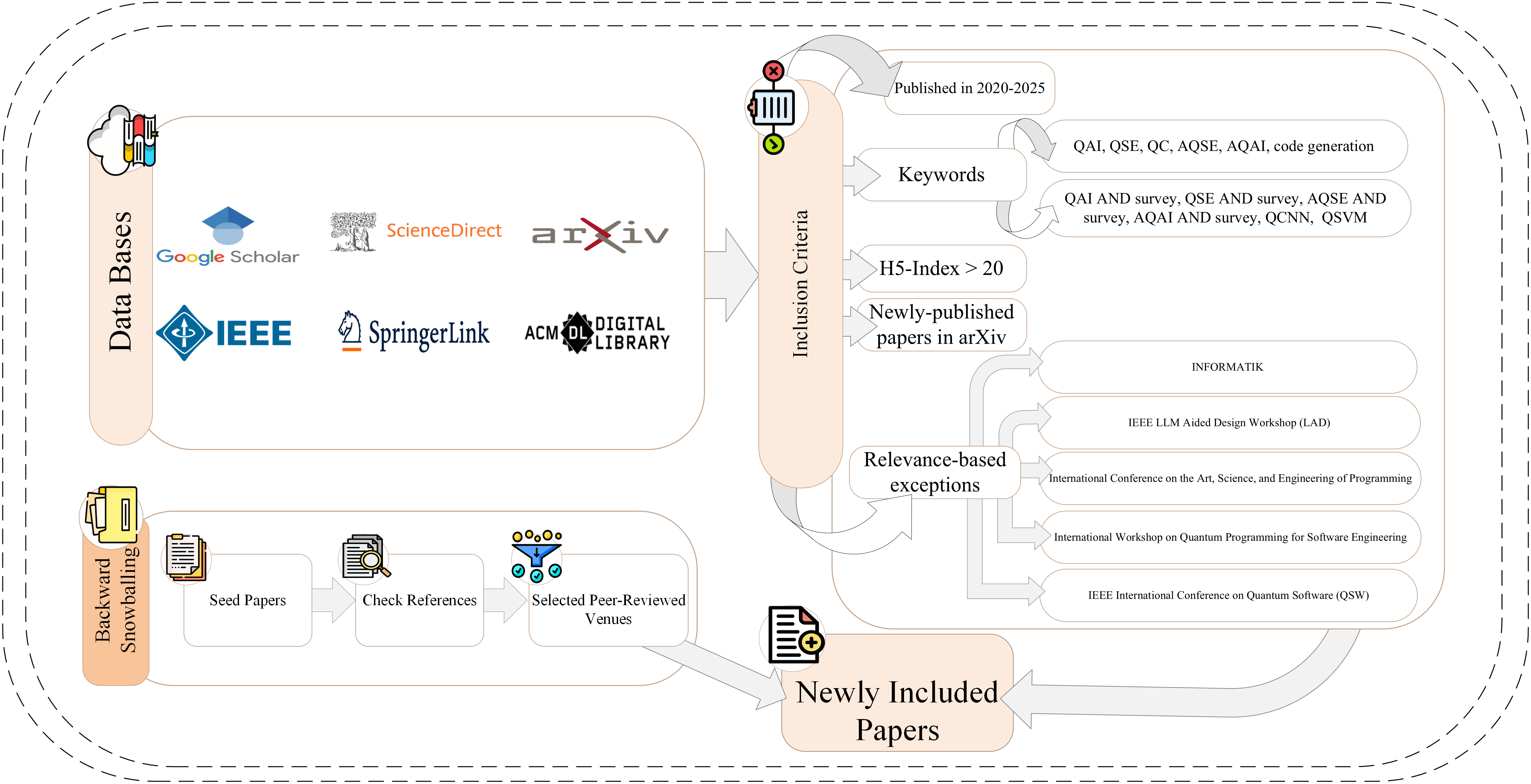}\caption{Overview of the Systematic Literature Review Methodology}
 \label{fig:research-method}
\end{figure*}
\section{Related Work}\label{sec:related-work}
There exist various surveys on topics related to QC, QAI, QSE, AQAI, and AQSE. However, most studies focused on one of these areas as a \textit{silo}. For instance, Gyongyosi and Imre \cite{GyongyosiImre2019} outlined the fundamental components essential for quantum computers, examined the necessary conditions for advancing large-scale quantum systems, and discussed the latest findings related to the physical realization of quantum devices, computational systems, and algorithms. Additionally, they explored several unresolved issues within the field. For example, a significant challenge remains in addressing the high error rates of quantum gates. Kusyk et al. \cite{Kusyk+2021} focused on methods that utilize AI and heuristics to address challenges related to Quantum Circuit Compilation (QCC). The authors categorized these approaches based on techniques employed, including various AI algorithms, such as genetic algorithms, genetic programming, and ant colony optimization, alongside heuristic methods, for example, greedy algorithms and optimization techniques related to dynamic programming and graph theory. The paper also discussed the performance of each QCC technique and evaluated its potential limitations. Yu and Zhao \cite{YuZhao2023} provided an extensive overview of Quantum Multi-Agent Reinforcement Learning (QMARL), a novel intersection of QC and multi-agent systems. It emphasized the transformative capabilities that QC brings to computational processes and explored the foundational concepts of multi-agent reinforcement learning (MARL). Their literature review investigated how integrating QC can enhance learning efficiency and decision-making. It synthesized existing literature, methodologies, and case studies that demonstrate the application of quantum algorithms within MARL frameworks. Additionally, it discussed the challenges and opportunities arising from quantum technologies in multi-agent contexts, such as entanglement and superposition, and their effects on agent coordination and learning dynamics. Practical implications are illustrated through examples in various fields, such as cybersecurity and finance, underscoring QMARL’s transformative potential. The paper concluded by outlining research gaps and proposing future directions, emphasizing the need for scalable quantum algorithms and quantum principles in agent collaboration strategies. Another study by Manan et al. \cite{Manan+2022} evaluated the advantages in speed and complexity that quantum computers may offer, providing an overview of the current understanding of ML applications in QC. Additionally, it evaluated the feasibility, performance improvements, and overall applicability of QC-ML algorithms. Further, in a review by Melnikov et al. \cite{Melnikov+2023}, the authors presented a dual-focused examination of methodologies that could enhance both quantum technology development and AI capabilities. The primary aim was to demonstrate how principles from physics can be translated into engineering solutions for machine learning using quantum software. They discussed quantum-enhanced algorithms that leverage QSE to optimize classical information processing, thereby improving essential ML techniques. The authors investigated how hybrid quantum-classical neural networks can enhance model generalization and accuracy while minimizing computational demands. They also illustrated how ML can mitigate the effects of errors on current Noisy Intermediate-Scale Quantum (NISQ) devices and facilitate the exploration of quantum advantages through automated studies of quantum walk processes on graphs. Furthermore, the paper assessed how ML can improve quantum hardware by addressing fundamental and applied physics challenges, including quantum tomography and photonics. Another review by Perez et al. \cite{Perez+2023} began with an introduction to both AI and QC, covering fundamental concepts and a timeline of major breakthroughs. It then shifted focus to existing research that illustrated the mutual benefits between QC and AI. The authors concluded by outlining prospective research directions within the emerging field of QAI. Additionally, a survey by Sarkar \cite{Sarkar2024} discussed the motivations for researching AQSE, offering a clear definition of the framework and insights into its necessary components for implementation. The authors also emphasized the important parallels between the quantum-classical interface of controllability and measurement and the interpretability and efficiency of AI models. This resemblance suggests that AQSE may experience similar successes and challenges as the Deep Learning (DL) revolution, ultimately necessitating a focus on hybrid approaches. Furthermore, Bacho et al. \cite{Bacho+2023} explored a critical question: Does the next generation of AI depend on QC? Their survey analyzed the limitations of classical digital computers, particularly those based on the Turing machine model, examining how these constraints influence the ability to solve specific problem categories, such as partial differential equations and optimization tasks, which in turn impact AI reliability. The findings revealed that many challenges present in traditional Turing machines also persist in QC models, including quantum circuits and quantum Turing machines. Finally, another review paper by Peral et al. \cite{Peral+2024} synthesized literature published over 2017-2023, identifying, analyzing, and categorizing various algorithms used in QML and their applications. 

\section{Research Methodology}\label{sec:methodology}
In this study, we deploy the Literature Review empirical research method (specifically the Systematic Review method) \cite{ACM-Empirical-Standards} to thoroughly and systematically survey and categorize the existing body of literature on QAI, QSE, AQSE, and AQAI topics. By systematically searching various databases and using a set of inclusion and exclusion criteria, we aim to improve the understanding of the interdisciplinary subject of this study in the scientific community. The search is divided into four primary categories as follows: (i) \textbf{Category A - QSE:} This category focuses on the intersection of QC and SE, investigating the fundamental definitions and various techniques in QSE. (ii) \textbf{Category B - QAI:} This category focuses on the intersection of QC and AI, examining quantum algorithms along with their applications in any sub-discipline of AI, including but not limited to machine learning. (iii) \textbf{Category C - AQSE:} Research in this category explores techniques for applying automation to any stage of the Software Development Life Cycle (SDLC) for quantum or hybrid quantum-classical software systems. (iv) \textbf{Category D - AQAI:} This category focuses on the integration of automation in developing, testing, deploying, and optimizing AI (including but not limited to machine learning) models, methods, and techniques, which will be deployed in AI-enabled software systems that leverage QC or hybrid quantum-classical computing.

To find relevant articles, we utilize several relevant academic databases, namely Google Scholar, IEEE Xplore, Science Direct, ACM Digital Library, and Springer Link. Additionally, arXiv is used to access recent novel papers, as it is an open-access platform for the pre-prints of scholarly articles. Despite not being peer-reviewed yet, many recent studies are only accessible there, making it a unique and valuable resource for a review in this fast-changing interdisciplinary area of research. Moreover, with the exception of arXiv, we limit our review to conference proceedings and journal articles from relevant and accredited peer-reviewed venues with an h5-index greater than 20. Due to the emerging nature of this field, some publication venues do not yet have an h5-index on Google Scholar. Based on our expertise, we decided to retain the following five venues despite the absence of an h5-index:
INFORMATIK, IEEE LLM Aided Design Workshop (LAD), International Conference on the Art, Science, and Engineering of Programming (Programming), IEEE International Conference on Quantum Software (QSW), International Workshop on Quantum Programming for Software Engineering.

Moreover, to capture the state of the art, we constrain our search to publications over the past 5 years and the current year (i.e., 2020-2025). Furthermore, the keywords we use to find the relevant research papers are as follows: \textit{QAI}, \textit{QSE}, \textit{QC}, \textit{AQSE}, \textit{AQAI}, \textit{code generation}, \textit{QAI AND survey}, {QSE AND survey}, \textit{AQSE AND survey}, \textit{AQAI AND survey}, \textit{QCNN}, and \textit{QSVM}, where \textit{AND} indicates a combination of two keywords. Further, another inclusion criteria we deploy besides the time of publishing and the stated keywords, which do not need to necessarily appear among the \textit{keywords} of the study explicitly, is whether the paper has directly addressed one of the four above-mentioned broad categories (i.e., QSE, QAI, AQSE, and AQAI). For instance, in Category B, which is QAI, the included papers must have covered both QC and AI. The papers that addressed only one of the categories, for example, only QC or only AI, will be excluded. Similarly, papers not published in the mentioned time frame or not published in one of the named venues shall be excluded. One exception is arXiv pre-prints, which are not excluded due to the reasons stated above.

\begin{figure}
\begin{center}
    \centering
    \includegraphics[width=0.75\columnwidth]{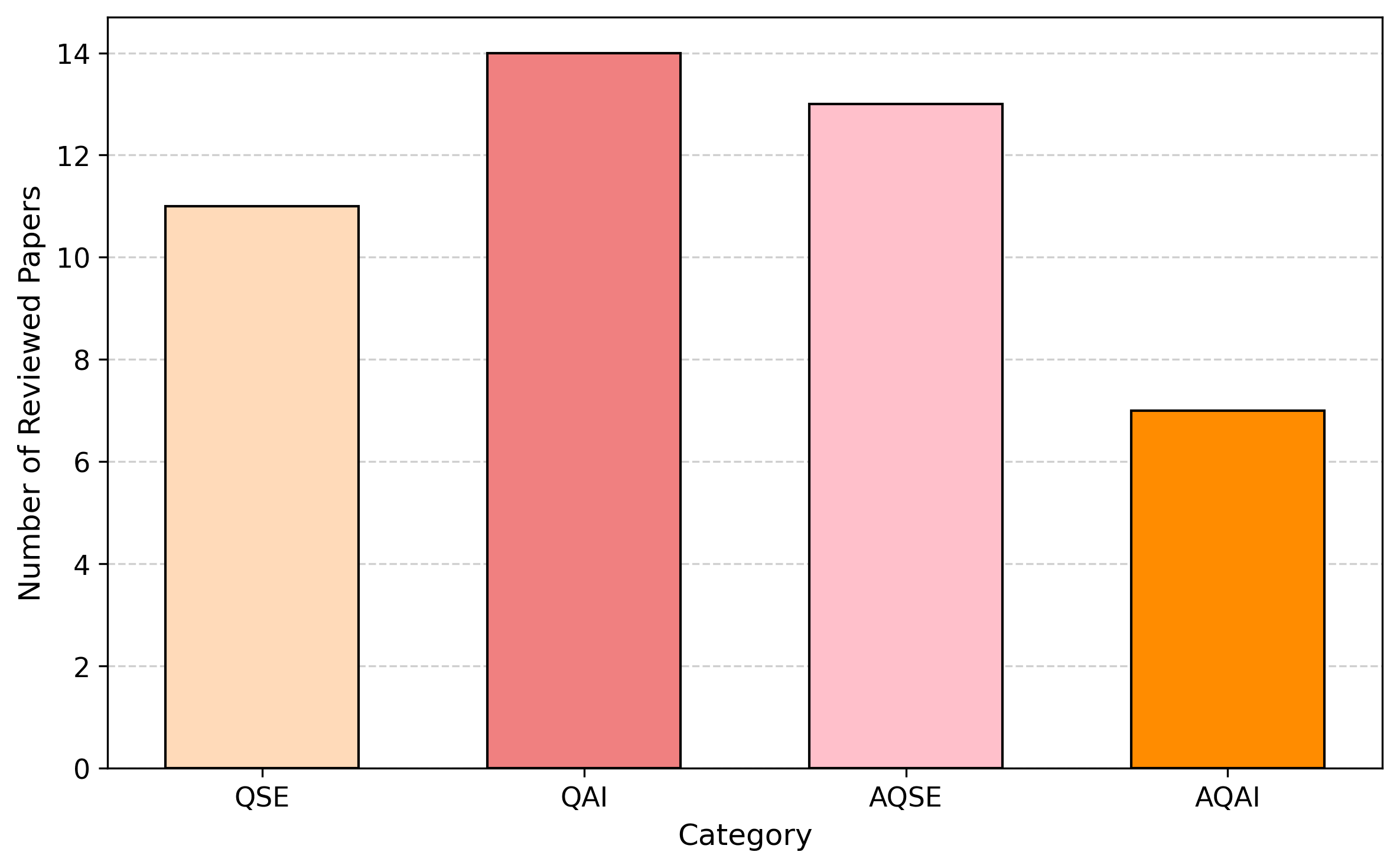}
    \caption{Distribution of Reviewed Literature by the chosen Category: QSE, QAI, AQSE, and AQAI}
   \label{fig:Distribution-reviewed-papers}
    \end{center}
\end{figure}

\begin{figure}
\begin{center}
    \centering
    \includegraphics[width=0.75\columnwidth]{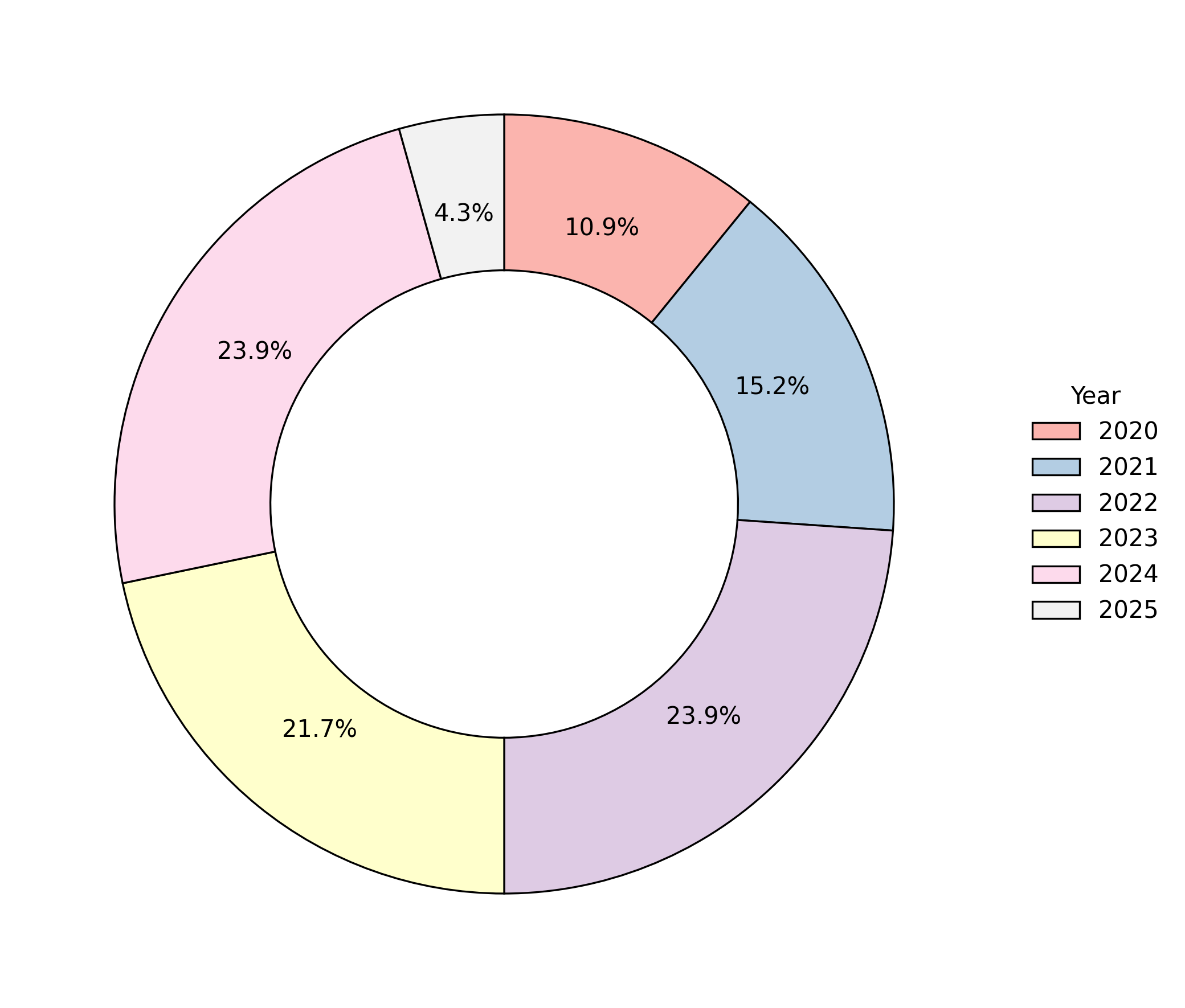}
    \caption{Distribution of Reviewed Literature by Year}
   \label{fig:Distribution-reviewed-papers-by-year}
    \end{center}
\end{figure}

\begin{figure}
\begin{center}
    \centering
    \includegraphics[width=0.75\columnwidth]{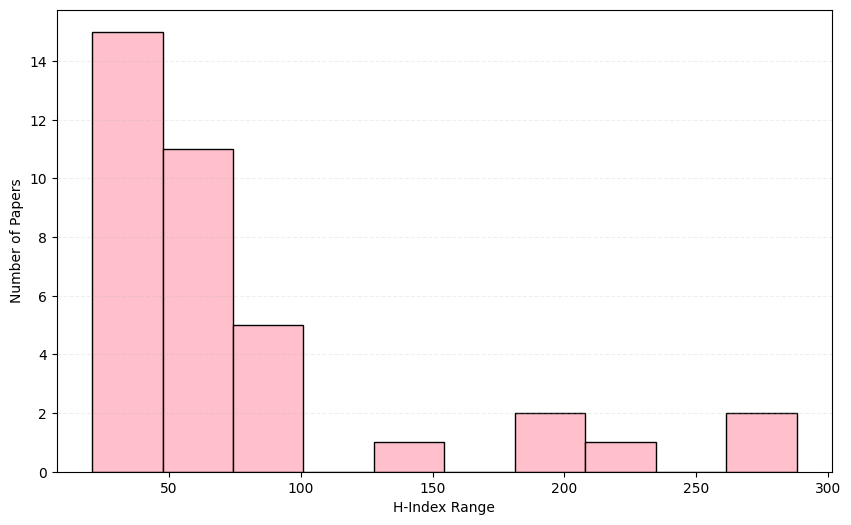}
    \caption{Number of Papers per H5-Index Range}
   \label{fig:Number-of-Papers-per-H-Index-Range}
    \end{center}
\end{figure}

In addition, we use backward snowballing of references \cite{Wohlin2014,JalaliWohlin2012}, which involves looking at the references of seed papers (the ones that are from good venues chosen by the authors) \cite{Sarkar2024, Mandal+2025, PerezPiattini2022} to find earlier relevant studies. The other exception to the venue inclusion/exclusion criteria is this case. The papers found through snowballing were published in (i) The Workshop on Quantum Software Engineering (QSE) in the IEEE/ACM International Conference on Software Engineering (ICSE) with the current Google Scholar h5-index of 84; (ii) IEEE/ACM International Conference on Automated Software Engineering (ASE) with the current Google Scholar h5-index of 54; (iii) The IEEE International Conference on Quantum Computing and Engineering (QCE) with the current Google Scholar h5-index of 28; (iv) The IEEE International Conference on Quantum Software (QSW), which is a relatively newer one with its 4th edition being held in 2025. The overall view of the proposed research methodology is depicted in Figure \ref{fig:research-method}. Also, the distribution of reviewed literature by category and year are available in Figure \ref{fig:Distribution-reviewed-papers} and \ref{fig:Distribution-reviewed-papers-by-year}, respectively. Lastly, Figure  \ref{fig:Number-of-Papers-per-H-Index-Range} illustrates the number of Papers per h5-Index range
\section{Quantum Software Engineering (QSE)}\label{sec:qse}
While QC offers significant advantages, realizing these benefits in practice requires robust software solutions. Therefore, it became essential to establish a new discipline known as \textit{Quantum Software Engineering (QSE)} \cite{Mandal+2025}. This field is critical to unlocking the full value of quantum applications and algorithms. The goal of classical Software Engineering (SE) is primarily employing sound principles, practices, and techniques to produce reliable industry-scale software systems based on their specifications within the limits of the allocated time and budget \cite{NaurRandell1969}. While classical SE is confined to software executing on classical (i.e., non-quantum) computers, software systems that are designed and developed for QC target quantum processors and quantum data (i.e., Qubits).

Jianjun Zhao's survey \cite{Zhao2020} defined QSE as the field related to utilizing sound, reliable, and economic engineering principles in the development process of software artifacts for software systems that should work efficiently on quantum hardware. Although the significance of the underlying hardware is critical, it is essential for both hardware and software development to evolve simultaneously. This parallel progression is vital to avoid a \lq{}quantum winter\rq{} \cite{Sarkar2024} (similar to what happened to AI twice in the past decades). This would, for example, be a scenario where expensive quantum devices exist without a clear understanding of their potential applications. The so-called Talavera Manifesto \cite{Piattini+2020} strove to collect a set of principles and qualities that they found necessary or reasonable for QSE.

\begin{figure*}
\begin{center}
    \centering
    \includegraphics[width=2\columnwidth]{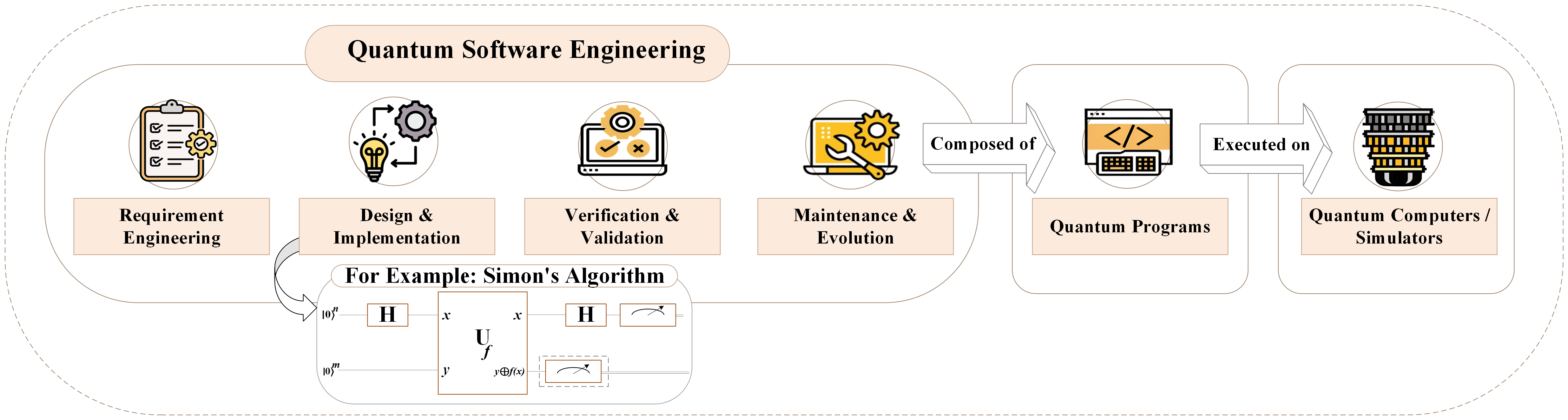}
    \caption{Overview of the QSE process \cite{Khan+2023,Ali+2022}}
   \label{fig:topics-qse-process}
    \end{center}
\end{figure*}

Figure \ref{fig:topics-qse-process} illustrates an overview of the QSE process.

Currently, the advancement of QC is largely driven by major tech companies. Google developed Quantum AI technologies using the Cirq \cite{cirq2023} and TensorFlow Quantum \cite{Broughton+2020} libraries and frameworks. These are based on the quantum circuits (i.e., gate-based) model of computation. In contrast, D-Wave focused on adiabatic quantum computers, leveraging the QMASM (Quantum Macro Assembler) programming language for optimization and machine learning tasks. IBM was a pioneer and offered QC as a Service (QCaaS) via its cloud platform with the open-source Qiskit programming language \cite{IBM-Qiskit, Javadi-Abhari+2024}. Furthermore, Xanadu Quantum Technologies developed photon-based hardware supported by full-stack open-source QC libraries, such as PennyLane \cite{Bergholm+2018}, allowing for scalability and robustness of QC at room temperature.

Cartiere \cite{Cartiere2021} presented a specification language designed to represent quantum computer operations through axiomatic definitions. This approach, termed formal quantum software engineering (F-QSE), utilized the same symbols and reasoning principles found in formal methods within software engineering, thereby enhancing the rigor of quantum software representation. One key limitation of their proposed approach is that it remained in its initial stage and was demonstrated using only a single example, the Deutsch algorithm, raising questions about its scalability and generalizability to more complex quantum algorithms and practical applications. Additionally, the work focused on basic quantum gates and a foundational algorithm, without addressing more advanced aspects of quantum computing. Wang et al. \cite{Wang+2020} proposed the quantum function-value binary expansion (qFBE) method, which facilitated the reversible implementation of algebraic functions. One limitation of this work is that the circuits, although modular, still needed many qubits and operations to get accurate results, which can be too demanding for current quantum devices. Furthermore, the circuits were only validated on a quantum simulator, leaving their practical feasibility on noisy intermediate-scale quantum (NISQ) hardware unconfirmed. Zhou et al. \cite{Zhou+2020} employed simulated annealing (SA) to optimize the initial mapping of input circuits. They used a heuristic cost function to iteratively apply the most effective SWAP gates until all quantum gates in the circuit could be executed. This work has several limitations, including slow performance on large circuits, unstable initial mappings from the SA algorithm, and the omission of key hardware constraints like crosstalk and decoherence. Additionally, the evaluation relies only on circuit size, without considering error rates or execution time. Another study by Zhao \cite{Zhao2023} introduced a preliminary catalog of code refactoring specifically designed for quantum programs. Their proposed refactoring tool is still under development and has not been publicly released or empirically validated, leaving its practical utility and effectiveness unproven. Bugs4Q \cite{Zhao+2021Bugs4Q}, a benchmark of real bugs for quantum programs, consisted of 36 verified Qiskit bugs across 4 widely used Qiskit programs. This benchmark included test cases and exploits, thus enhancing the reproducibility of empirical studies and enabling effective comparisons of various tools for analyzing and testing Qiskit. Although Bugs4Q provides a valuable benchmark, a key drawback is that it includes only 36 reproducible bugs limited to Qiskit components, which may not fully reflect the diversity of quantum programming errors. Yue et al. \cite{Yue+2023} specifically focused on Quantum Software Requirements Engineering (QSRE) to address the challenge of ensuring accurate and reliable software requirements in quantum systems. This paper presented preliminary ideas and discussions rather than a fully developed methodology or framework. Future work should focus on formalizing the approach and evaluating it through practical applications and empirical studies.

Overall, the trend in the rather new field of QSE seems to comprise expanding and extending various techniques, methods, and practices in different stages or phases of the software process from the classical world to the quantum and hybrid quantum-classical settings. Table \ref{tab:literature-classification} summarizes the reviewed literature in the QSE category.

\section{Quantum Artificial Intelligence (QAI)}\label{sec:qai}
The integration of AI (particularly, its sub-discipline machine learning) with QC enhances the extraordinary processing capabilities of quantum systems \cite{Kottahachchi+2025}. The interplay of AI and QC can have mutual benefits for the advances of both fields. This emerging field, is also referred to as quantum intelligence. A significant part of it is about the sub-discipline of AI, machine learning, which is currently the hottest and most widely used area of AI across the industry. This sub-field of QAI is known as Quantum Machine Learning (QML). Research in QML aims to leverage quantum characteristics and parallelism to boost performance in areas such as pattern recognition and optimization. By harnessing the strengths of both AI and QC, this combination holds the promise of transforming various industries. QC's ability to handle extensive datasets and execute intricate calculations at remarkable speeds positions it as a powerful asset for sophisticated machine learning and data analytics tasks. At the core of quantum algorithms for machine learning resides a structure rooted in classical theories, enhanced by quantum phenomena such as superposition and entanglement.

\begin{figure}[h]
\centering
\includegraphics[width=0.5\columnwidth]{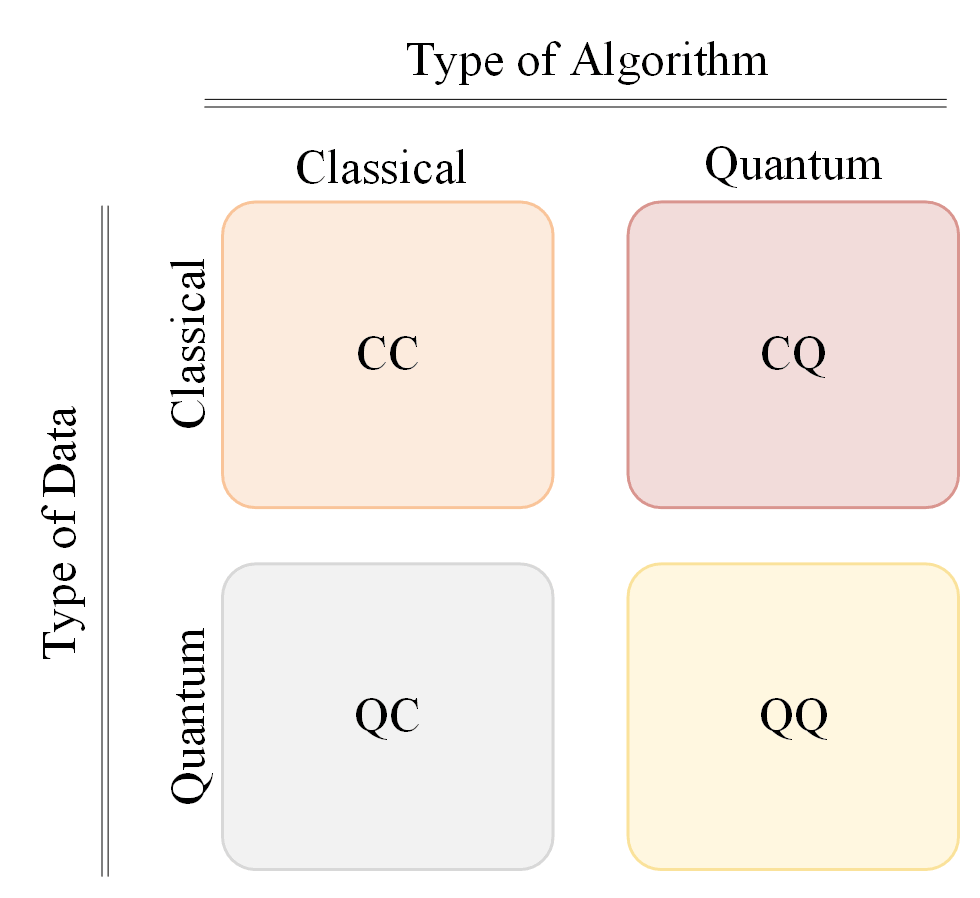}\caption{A typology of QML from the data and algorithms perspectives \cite{Aimeur+2006}}
 \label{fig:topics-typology-qc-ml-integration}
\end{figure}

A typology suggested by Aimeur et al. \cite{Aimeur+2006} categorizes QML into four distinct scenarios based on the data and algorithms. This is depicted in Figure \ref{fig:topics-typology-qc-ml-integration}. This classification depends on whether the data are in qubit form (Q) or classical bits (C), and whether the algorithms are quantum by design (Q), meaning that they can benefit from the characteristics of quantum processors, or they are classical algorithms (C) \cite{Aimeur+2006}. In fact, the CC scenario involves classical data processed with classical algorithms, whereas QC is the scenario in which data are in qubits form but the algorithms are classical; CQ uses classical data with quantum algorithms; and the QQ scenario possesses both data and algorithms in quantum form.

Some of the challenges of QC, especially when it comes to QAI are as follows \cite{Fernandez+2022}: (i) The slow data throughput, especially for reading the input data, may outweigh the computational benefits of quantum algorithms, limiting speed-up. (ii) Translating quantum algorithm outputs into binary format requires learning an exponential number of bits, making some QAI applications impractical. (iii) Rapid fluctuations in quantum states can lead to information loss, complicating quantum error correction and threatening the validity of analyses in QAI. (iv) There are few benchmarks for assessing quantum algorithm performance, and a solid framework for comparing QAI to classical AI is lacking. (v) Estimating the number of quantum gates needed for specific algorithms remains challenging, despite theoretical advantages suggested for QAI.

Table \ref{tab:literature-classification} presents the reviewed literature in the QAI category.

In the following, we elaborate on two broad sub-categories of QML, namely supervised and unsupervised. These are parallels to the classical supervised and unsupervised machine learning categories.

\subsubsection{Quantum Supervised Learning}
Quantum supervised learning is an emerging field that combines principles from QC with classical supervised machine learning methods \cite{Macaluso+2020}. Following the quantum circuit (gate-based) model of computation, in a typical quantum supervised learning classifier, classical data with associated class labels or target values are fed into a quantum circuit \cite{BhowmikThaplival2024}. This circuit generates a probability distribution across various classes. After each iteration, the output is measured to yield a probability distribution over different quantum states \cite{BhowmikThaplival2024}. Among the primary approaches in quantum supervised learning, Quantum Neural Networks (QNNs) are currently the most prominent family of methods. QNNs can be further extended into Hybrid Quantum-Classical Neural Networks (HQCNNs).

Javaria et al. \cite{Javaria+2022} introduced a significant model that employed specific learning parameters to generate synthetic Computed Tomography (CT) scan images resembling actual scans. This work engaged QNNs in analyzing and classifying medical images related to COVID-19. Their framework classified only the presence or absence of COVID-19. In future work, the system could be extended to classify the severity levels of infection (mild, moderate, and severe), which would be highly valuable in clinical decision-making and improving patient survival rates. In another paper, Matic et al. \cite{Matic+2022} proposed a hybrid quantum-classical convolutional neural network architecture for detecting breast cancer in 2D ultrasound images and classifying organs in axial CT scans. Their research represented a significant advancement in 3D medical imaging, demonstrating the capabilities of quantum-classical networks for the first time on such datasets. However, this study does not yet show a clear advantage of QCCNNs over classical CNNs in training or generalization, especially with limited data. While the quantum model uses fewer parameters, further research is needed to explore the impact of circuit design, filter count, and encoding strategies, particularly for high-resolution 3D medical data and hybrid architectures. Additionally, Gohel et al. \cite{Gohel+2022} developed a QML model for classifying organs using X-ray datasets of the hand, ankle, knee, and chest. This model deployed a deep quanvolutional neural network, which combined quantum encoding for input data with classical neural network processing. In their study, the quantum component relies on a basic circuit design using only 4 qubits and Ry rotation gates. While suitable for initial experimentation, this limited setup constrains the circuit’s expressiveness and raises concerns about its ability to scale and capture the complexity required for real-world medical image analysis. 

From our literature review, it becomes evident that a significant portion of the research work in the emerging field of quantum supervised learning focused on the quantum or hybrid quantum-classical variants of classical supervised machine learning algorithms, most importantly QCNN and QSVM.

\subsubsection{Quantum Unsupervised Learning}
Quantum unsupervised learning prominently features dimensionality reduction and clustering techniques \cite{UllahGarcia2024}. One notable application of quantum clustering methods is in enhancing privacy since quantum unsupervised learning supports reduced data access frequency and volume due to the efficiency of the employed quantum algorithms.

Neukart et al. \cite{Neukart+2018} proposed Quantum-Assisted Clustering Analysis (QACA), which leveraged the topological properties of the D-Wave 2000Q quantum processor. By utilizing a Quantum Processing Unit (QPU), they framed optimization, sampling, and clustering tasks as Quadratic Unconstrained Binary Optimization (QUBO) problems. Their experiments showed that QACA achieved an accuracy comparable to classical clustering algorithms, although the results could vary depending on the clustering methods used. One future direction of this research is to further leverage the spatial layout of activated qubits on the QPU to construct a comprehensive feature map, potentially enhancing clustering quality. Further, Wang et al. \cite{Wang+2023} presented a classical-to-quantum transfer learning framework that utilized a large-scale unsupervised pre-trained model to illustrate the effectiveness of quantum transfer learning for synthetic speech detection. They utilized the pre-trained WavLM-Large model to derive feature maps from speech signals, generating low-dimensional embedding vectors through classical network components. This approach involved jointly fine-tuning both the pre-trained model and the classical network components using a variational quantum circuit (VQC). A limitation of this study is that, while it improved the classification accuracy of in-set real samples, it resulted in performance degradation for out-of-set real samples. This suggests that the quantum circuits are prone to overfitting, particularly when the number of training samples for the real class is limited and the quantum model uses only a small number of qubits.

\section{Automated Quantum Software Engineering (AQSE)}\label{sec:aqse}
As mentioned earlier, QC has the potential to fundamentally change how we address complex issues that classical computers struggle to solve. However, its distinct features and inherent challenges require a specialized approach to developing and deploying software. The field of QSE can be quite challenging, mainly because quantum logic often operates in ways that are not intuitive. As quantum systems grow in size and complexity, traditional programming techniques, such as quantum assembly languages and qubit-level reasoning, become less practical \cite{Murillo+2024,Dwivedi+2024}. This is where AQSE comes into play. AQSE can help simplify the development process, making it more accessible, speeding up production, minimizing errors, and enhancing the overall efficiency of quantum software. AQSE involves applying automation techniques to the software process, from design and implementation to testing and optimization of quantum software. Notable applications of AQSE include quantum code generation, automated quantum error correction, and automated verification and testing of quantum software.

The reviewed literature in the AQSE category are itemized in Table \ref{tab:literature-classification}.

Despite the progress made in AQSE, the challenges below still remain in the field \cite{Asif+2025, Saxena+2024}: (i) Quantum programming lacks large-scale, high-quality datasets, which hinders the development and training of robust models. (ii) Limited access to quantum hardware - typically available only through a few cloud providers -can slow down both manual and automated development and testing processes. (iii) Integrating classical and quantum processing requires the code generation technology to understand and generate code across hybrid paradigms, posing a significant challenge. (iv) Quantum systems are inherently prone to errors and noise, making it difficult to ensure reliability, even with extensive testing.

In what follows, we review automation techniques in QSE, categorizing them as either AI-based or non-AI-based quantum software artifact generation techniques.

\subsection{AI-Based Approaches} 
Large Language Models (LLMs) have become vital assets in the realm of software development, enhancing the coding process and significantly lowering the time and effort needed to build applications \cite{Dupuis+2024,SiavashMoin2025, SiavashMoin2025-models}. Saxena et al. \cite{Saxena+2024} introduced a Quantum DevOps framework designed to enhance continuous integration and continuous deployment (CI/CD) processes tailored for quantum algorithms. This framework aimed to streamline the integration and deployment of quantum applications. Its key features included the utilization of LLMs for automating the generation of microservice code, deployment on Kubernetes (K8s) clusters for scalable infrastructure management, and a centralized registry for quantum artifacts to ensure effective version control and resource management. They identified several areas for future improvement of their research, including advanced error correction and noise mitigation, hybrid quantum-classical workflows, scalable testing frameworks for quantum applications, integration of the Quantum DevOps toolchain, management of the quantum software lifecycle, and the development of educational and training programs. In another study by Dupuis et al. \cite{Dupuis+2024}, the authors presented specialized LLMs designed to serve as code assistants for users of the IBM Qiskit SDK. As part of future work, the authors of this research anticipated the need for additional capabilities such as code explanation, translation between different libraries or versions, automatic test generation, and code repair to address the evolving demands of QSE. Furthermore, Asif et al. \cite{Asif+2025} introduced a high-quality dataset of 3,347 PennyLane-specific \cite{Bergholm+2018} code samples and their contextual descriptions aimed at training an LLM-based quantum code assistance. While this research provided a valuable contribution by focusing on the PennyLane framework, its scope remains limited to a single quantum computing platform. To enhance the generalizability and applicability of the approach, future work should consider expanding the dataset and methodology to include other frameworks such as IBM's Qiskit and Google's Cirq. In MDE4QAI \cite{Moin+2022-MDE4QAI}, the authors proposed a vision for using the Model-Driven Software Engineering (MDSE) paradigm to support software developers and data scientists in transitioning to QC and QAI, as well as to hybrid quantum-classical software systems, including such systems that are AI-enabled, building on the use of MDSE in the automated production of classical AI-enabled software systems and services \cite{Moin+2022ML-Quadrat}. As Moin et al. mentioned, future research should improve modeling tools to support the unique needs of new quantum processors with different architectures, making it easier to use them in future smart systems. Additionally, an approach for implementing model-driven Quantum Federated Learning (QFL) was presented by Moin et al. \cite{Moin+2023-QFL}. While their current work suggested extending existing domain-specific model-driven engineering tools for machine learning enabled systems, developing a working prototype and evaluating its performance would help demonstrate the practical value of the approach. Also, Polat et al. \cite{Polat+2024} proposed a framework that integrated the gate-based (quantum circuits) and quantum annealing (adiabatic) models of computation for QC using MDSE. A potential direction for future work involves expanding the Platform-Independent Model (PIM) to accommodate new quantum technologies, along with incorporating Domain-Specific Languages (DSLs) to simplify the development of quantum algorithms across different platforms. 

Recent trends in AQSE indicate a growing emphasis on code generation utilizing LLMs. This shift reflects a broader trend concerning the increasing interest in leveraging advanced machine learning techniques to automate and enhance the software development process. Note that LLMs are also considerably transforming the code generation methods in MDSE \cite{DiRocco+2025}.

\subsection{Non-AI-Based Approaches}
In contrast to AI-driven methodologies, some non-AI-based approaches emphasize using conventional software engineering techniques to aid in quantum software development. These methods mainly utilize formal modeling, classic non-AI-based approaches to code generation in MDSE, and secure code transformation strategies to optimize the quantum software development lifecycle. Lavrijsen et al. \cite{Lavrijsen+2020} introduced a set of optimizers specifically designed for application on NISQ devices. Their study evaluated the performance and effectiveness of various optimizers through a case study involving Variational Quantum Eigensolver (VQE), which operates as a hybrid algorithm. In their framework, a classical minimization step directed subsequent evaluations on the quantum processor. The findings suggested that optimizers tailored for specific tasks are essential for achieving reliable scientific outcomes with NISQ hardware, and this requirement was expected to persist even as advancements were made toward fault-tolerant quantum circuits. Future work could look into how different ansatz designs affect optimizer performance. This study mainly uses the unitary coupled cluster ansatz, but trying other or more tailored ansatz might improve results and help the optimizer work better in noisy conditions. Additionally, Jim{\'e}nez-Navajas et al. \cite{Jimenez+2025} proposed a technique for generating quantum code from extended UML design models. This method involved a series of model-to-text transformations, implemented using the Epsilon generation language, to produce both Python and Qiskit code, effectively merging classical and quantum programming. The approach was validated through a multi-case study involving seven hybrid software systems modeled in UML, demonstrating its effectiveness and efficiency. The paper suggested that the software modernization process for hybrid systems could advance by addressing the forward engineering phase, thereby promoting the broader adoption of quantum software through MDSE. However, a notable limitation of this method was its emphasis on generating code for circuits fully represented in activity diagrams, excluding modeling abstractions for circuits that would adhere to specific design patterns or algorithm families, such as the Quantum Fourier Transform (QFT) or VQE.

\section{Automated Quantum Artificial Intelligence (AQAI)}\label{sec:aqai}
Although AI's scope reaches far beyond machine learning, our discussion in this section is bounded to automated approaches to Quantum Machine Learning (QML) practices. The optimization of classical machine learning algorithms depends on the choice of specific model families and hyperparameters that must be meticulously adjusted, often through trial and error, to learn from data effectively in order to become capable of inference on unseen instances. There exist automated approaches to this optimization task; A field known as Automated Machine Learning (AutoML). Note that this is expected to become even more beneficial to QML as quantum adaptations of machine learning algorithms introduce numerous additional parameters and circuit designs that require careful tuning \cite{Gomez+2022}. Automation can simplify the process of generating and training quantum classifiers \cite{Altares+2024}. To enhance the understanding of how AQAI, specifically AQML, operates, several relevant studies are reviewed below. Additional references are listed together with these under the AQAI category in Table \ref{tab:literature-classification}.

Altares-L{\'o}pez et al. \cite{Altares+2024} proposed a method for automatically creating and training quantum classifiers specifically tailored to grayscale images, utilizing multiobjective genetic algorithms. Although their current representation may not be sufficiently extensive for more complex classification problems, the proposed methodology is inherently scalable. Future work could explore increasing the resolution of the encoding scheme by adding more bits per individual, thereby enabling a larger set of gates and angles to be represented. Moreover, Roth et al. \cite{Roth+2025} introduced an innovative framework that applied AutoML to QML. This framework offered a modular and cohesive programming interface that simplified the creation of QML pipelines. The AutoQML framework utilized the QML library sQUlearn to enable a range of QML algorithms. Evaluations conducted across four industrial applications showed that AutoQML could produce high-performing QML pipelines. Future extensions of this framework could include support for unsupervised learning tasks such as clustering and exploration of reinforcement learning methods, expanding its applicability across a broader range of machine learning domains. Additionally, strategies to reduce the computational cost of QNNs would help ensure their more frequent inclusion in the pipeline search, improving the diversity of QML models explored by the framework. Additionally, a framework named AutoAnsatz was presented by Koike-Akino et al. \cite{Koike+2022} to support the design of QNN architectures. Most QNN models utilized specific quantum circuit templates known as \textit{ansatz}, which still required careful adjustment of hyperparameters, including the number of qubits and the number of entangling layers. This study presented an initial proof-of-concept for applying AutoQML to design QNN ansatz for Wi-Fi sensing tasks. The approach was evaluated using an in-house Wi-Fi testbed and demonstrated that small-scale QNNs could achieve performance comparable to large-scale DNNs in human pose recognition. However, validation on real quantum hardware was not conducted and is planned for future work. Many open challenges remain, including the scalability and generalization of the method in more complex sensing scenarios. Further, Incudini et al. \cite{Incudini+2024} introduced an innovative method to tackle the complex issue of choosing the right quantum kernel for a given task. They developed an algorithm that automated the discovery of quantum kernels, functioning without any prior knowledge about the problem at hand and relying only on a set of observations. This algorithm generated candidate solutions through heuristic optimization. In their experiments, the approach successfully identified quantum kernels that either matched or exceeded the performance of the best existing methods. While this study demonstrates the potential of their proposed approach, it only explored a limited subset of configurations. Future work could investigate the impact of individual criteria, compare alternative optimization algorithms, including those designed for high-dimensional search spaces, and apply the method to related anomaly detection tasks to further assess the practicality of quantum kernels.

Altares-L{\'o}pez et al. \cite{Altares+2021Automatic} in another study addressed the challenge of supervised learning through quantum feature maps optimized by a genetic algorithm. The method aimed to enhance both accuracy and generalization while reducing circuit size. The authors evaluated this technique using both synthetic and real-world benchmarks, achieving impressive accuracy across all tested problems. Future research could extend the current gene encoding scheme by incorporating a broader set of local and entangling gates, increasing the precision of parameterization, and exploring different gate orderings.
 Another study by Gomez et al. \cite{Gomez+2022} proposed a hybrid cloud architecture that combined classical computing and QC, facilitating parallel hyperparameter exploration and model training. Their sample configuration file captured hundreds of potential quantum Generative Adversarial Networks (qGANs). Future work on this study could focus on improving the model selection process by integrating advanced optimization techniques, such as Bayesian optimization, to enable more efficient exploration beyond brute-force search. Also, enhancing the framework to support automated quantum circuit design, including dynamic ansatz generation and gate selection, would further increase flexibility.  Lastly, the AQMLator platform was introduced as an AQML solution designed to automatically propose and train the quantum layers of a machine learning model with minimal user input \cite{Rybotycki+2024}. This open-source platform could run on both simulators and real quantum devices via Qiskit. It autonomously suggested QML models based solely on the provided data and task from the user. Built with standard QML libraries, AQMLator could facilitate smooth integration of the resulting models into existing QML workflows. Additionally, it was thoroughly documented, making it user-friendly and easily customizable to fit individual requirements. Although AQMLator is functional in its current form, several enhancements could broaden its usability and effectiveness. Future development could include adding more default layers, data embedding methods, and models to reduce the user’s manual setup. Incorporating advanced reinforcement learning-based quantum architecture search would enable a more thorough exploration of the circuit space. Additionally, streamlining the integration of new evaluation metrics and enabling native multi-objective optimization with Optuna would allow users to better balance trade-offs, such as cost versus model quality.

While AQAI significantly simplifies the optimization process through automation, it also faces challenges that can complicate its implementation: (i) There is currently a shortage of professionals skilled in both QC and Automated AI, which impedes advancement in the field. (ii) Automated model selection and hyperparameter tuning in the quantum realm is quite tricky because it involves unique quantum-specific factors, such as circuit depth and qubit connectivity.
\begin{table*}
\caption{The reviewed literature that directly related to one of the four categories, classified according to various criteria}
\begin{center}
    \centering
\resizebox{1\textwidth}{!}{ 
\scriptsize
\begin{tabular}{|c|l|>{\centering\arraybackslash}p{0.25\linewidth}|>{\centering\arraybackslash}p{0.26\linewidth}|>{\centering\arraybackslash}p{0.3\linewidth}|>{\centering\arraybackslash}p{0.25\linewidth}|c|} 
\hline 
           \textbf{Ref} & \textbf{Year} & \textbf{Dataset/Input} & \textbf{Method/Framework} & \textbf{Evaluation Metrics} & \textbf{Tasks} & \textbf{Category} \\ 
\hline 
           \cite{Tavassoli+2024}&   2024& N/A&  Quantum Computing as a Service for Hybrid
Classical-Quantum Software Development (QCSHQD)&  N/A&  Quantum software development& \multirow{12}{*}{\raisebox{-15em}{\rotatebox{90}{\fontsize{10}{8}\selectfont \textbf{Category A: QSE}}}}\\ \cline{1-6} \cite{Fortunato+2024}&   2024& Generated by QuraTest&  Gate Branch Coverage (GBC)&   GBC score: 100\%&  Quantum software testing&  \\ \cline{1-6} \cite{Yue+2023}&   2023& N/A&  QSRE&  N/A&  Ideas on how Requirements engineering (RE) for quantum software will differ from the classical counterpart&  \\ \cline{1-6}\cite{Zhao2023}&   2023& Main and performquantumsimulation operations&  Refactoring Quantum Programs in Q\#&  N/A&   Facilitating the effective restructuring of quantum
programs in Q\#&  \\ \cline{1-6} \cite{Perez+2022}&   2022& 9 D-Wave programs&  
Reverse engineering of Hamiltonian expression&  KDM g. T(s): 0.0080-0.2187 H.g. T(s): 0.0032-0.3738&  Software modernization&  \\ \cline{1-6}\cite{PontolilloMousavi2022}&   2022& Simulated Input&   Multi-lingual benchmark&   Mutation 
testing&  Property-based testing of
quantum programs&  \\ \cline{1-6}
 \cite{Cartiere2021}& 2021& N/A& F-QSE& N/A& Offering more intuitive approach to thinking and writing about QC systems&\\ \cline{1-6}\cite{Zhao+2021Bugs4Q}& 2021& Produced buggy behaviors& Bugs4Q& Manual checks, dynamic validation through version control, and thorough documentation& Bug classification&\\ \cline{1-6}\cite{Wang+2020}&   2020& Logarithmic, exponential,
trigonometric and inverse trigonometric functions &  New evaluation method called qFBE&  qFBE&  Evaluating transcendental functions&  \\ \cline{1-6} \cite{Zhou+2020}&   2020& Logical quantum circuit&  SA-based algorithm&  Percentage of the number of added gates in the new algorithm to the compared one (Comparison): 0-92.40\%, Running time: 0-1059.06 (s)&  Quantum circuit transformation&  \\ \cline{1-6}\cite{KrugerMauerer2020}&   2020& Simulated Dataset&  Transforming Boolean satisfiability problem (SAT) to a quantum
annealer&  Accuracy  (ACC) difference: - 0.2-0.6
Scalability&  Software augmentation with quantum components&  \\ &   & &  &  &  &  \\  \hline
 \cite{Ren+2024}& 2024& Simulated Dataset& QFDSA& ACC: 97.70-99.10\%, Precision (P): 96.91-99.81\%, F1: 95.60-98.81\%, (S): 98.06-99.88\%&Smart grid dynamic
security assessment& \\ \cline{1-6} \cite{Nadim+2024}& 2024& 20 Software Defect Datasets& Variational Quantum Classifier (VQC), Quantum Support Vector Classifier (QSVC), Pegasos Quantum Support Vector Classifier (PQSVC)& F1: 0.55-0.76&Software defect prediction&  \multirow{14}{*}{\raisebox{-20em}{\rotatebox{90}{\fontsize{10}{8}\selectfont \textbf{Category B: QAI}}}}\\ \cline{1-6}\cite{Chen+2023}& 2023& Anonymous Breast
Cancer Dataset& QCNN, hybrid
Quantum-classical
QCNN& ACC: 78–93\%&Breast cancer classification&  \\ \cline{1-6} \cite{Tripathi+2023}& 2023& Simulated Dataset& QML binary classification model& P: 0.258-0.941, Recall (R): 0.501-0.940, F1: 0.341-0.939, ACC: 0.51-0.94&Cyber threat detection&  \\ \cline{1-6} \cite{Salari+2023}& 2023& Interaction free Imaging Dataset, Ghost Imaging Dataset& QPCA, QICA& Processing time: 115.5-6781.6, Complexity: $NLog N$&Face recognition&  \\ \cline{1-6}\cite{Bhavsar+2023}& 2023& Simulated Dataset& VQC and
PQSVC& P: 88.74-97.96 \%, R: 85.73-97.96, \% F1: 85.29-97.94 \%, ACC: 85.73-98.11\%&Hazardous asteroids classification&  \\ \cline{1-6} \cite{Caivano+2023}& 2023& MQTTset Dataset& QBoost, PegaSOSQSVM, VQC& ACC: 0.61-0.82, P: 0.68-0.87, R: 0.61-0.82, F1: 0.54-0.82& Intrusion Detection&  \\ \cline{1-6}\cite{Caivano+2023extending}& 2023& Car-Hacking Dataset& Secure Quantum Automotive Development and Engineering (SeQuADE)& ACC: 0.9941, P: 0.9871-1.0, R: 0.8291-1.0, F1: 0.9065-1.0&Controller Area Network (CAN) Attack Identification&  \\ & & & & & &  \\ \cline{1-6}\cite{Grossi+2022
}& 2022& European cross-border payment transactions dataset& QSVM& ACC: 0.55-0.78, AUC: 0.74-0.81&Financial crime
prevention&  \\ \cline{1-6} \cite{Ciaramella+2022}& 2022&  Android malware and legitimate applications Datasets&  LeNet, AlexNet, a
CNN model designed by authors, VGG16, HQNN& Loss: 0.15-1.87, ACC: 0.25-0.95, P: 0-0.96, R: 0-0.95, F1: 0-0.95, AUC: 0.62-0.99&Mobile malware detection&  \\ \cline{1-6} \cite{Javaria+2022}& 2022& POF Hospital, UCSD-AI4H dataset& Conditional Adversarial Network (CGAN), Quanvolutional neural network& ACC: 0.94-0.96, R: 0.94-0.95, F1: 0.94-0.96, P: 0.94-0.96&COVID-19 classification&  \\ & & & & &&  \\ \cline{1-6} \cite{Matic+2022}& 2022& MedMNIST, Lung-nodule Datasets& QCCNN& ACC: 0.4-1, Loss: 0-1.7&Radiological image classification&  \\ \cline{1-6}\cite{Gohel+2022}& 2022& Organ X-Ray Image Dataset& HQCNN& ACC: 92.5\%, R: 90\%, F1: 90\%&  Organ classification&  \\ \cline{1-6} \cite{Schuld+2020}& 2020& CANCER, SONAR, WINE, SEMEIONand MNIST256
Datasets& Variational Quantum algorithm& Train and Test error:  0.000-0.058&Creation of a quantum classifier&  \\ & & & & &&  \\ \hline
  \cite{Jimenez+2025}& 2025& 7 hybrid software systems& MDE& P: 65-100\%, R: 80-100\%, F1: 74-100\%&Code generation for classical-quantum software systems modeled in UML& \multirow{13}{*}{\raisebox{-18em}{\rotatebox{90}{\fontsize{10}{8}\selectfont \textbf{Category C: AQSE}}}}\\ \cline{1-6} \cite{Polat+2024}& 2024& Full-Entangled Ising Model, Linear-Entangled Ising Model, Circular-Entangled Ising Model & MDE4QP& Ground state energy calculation& Integrating quantum annealing and gate-based quantum devices for ground state energy solutions in multi-partite quantum systems&  \\ \cline{1-6} \cite{Dupuis+2024}& 2024& GRANITE-20B-CODE-QK, DEEPSEEK-CODER-33B-BASE, GRANITE-20B-CODE-QK, DEEPSEEK-CODER-33B-BASE& LLM& Qiskit HumanEval (QHE): 20.79-46.53\%&QC code generation&  \\ \cline{1-6} \cite{Saxena+2024}& 2024& N/A& CI/CD, LLM, K8s& N/A&Automated testing and deployment, microservice code generation&  \\ \cline{1-6} \cite{AragonesOriol2024}& 2024& Simulated Dataset& C4Q: A Chatbot for Quantum& Loss: 0.0001-0.1, ACC: 0.99-1.00&&  \\ \cline{1-6} \cite{Moin+2023-QFL}& 2023& N/A& Domain-specific MDE& N/A&Model-driven QFL&  \\ \cline{1-6}\cite{Moin+2022-MDE4QAI}& 2022& N/A& MDE4QC, MDE4QAI& N/A&QML for the IoT and
smart CPS applications&  \\ \cline{1-6} \cite{Fortunato+2022}& 2022& 11
real quantum programs& QMutPy& Average mutation score: 57.7\%&Mutation testing of quantum programs&  \\ \cline{1-6} \cite{Ali+2021}& 2021& Ent, Swap, RCR, Inc, Dec& Quito (QUantum InpuT
Output coverage)& $\text{ms}_{\text{TOTAL}}$: 72.8-95\%
$\text{at}_{\text{TOTAL}}$: 2001-160990&  Quantum software testing&  \\ \cline{1-6} \cite{Wang+2021}& 2021& QRAM, QRAMmut& Quito& Wrong Output Oracle (WOO): 0
Output Probability Oracle (OPO): 0-50
Time: 81-1795&Test generation, execution, and assessment&  \\ \cline{1-6} \cite{Mendiluze+2021}& 2021& IQFT, QRAM, BV, CE& Muskit (Mutation Analysis Tool)& WOO: 0-99.26\%
OPO: 0-100&Quantum software testing&  \\ \cline{1-6} \cite{Wang+2021qdiff}& 2021& X gate, Deutsch-Jozsa, Bernstein-Vazira, Grover, Variational-
Quantum-
Eigensolver (VQE), Quantum
Approximate
Optimization
Algorithm (QAOA)& QDIFF& Speed up
differential testing by 66\%&Differential testing for Quantum Software
Stacks (QSS)&  \\ \cline{1-6} \cite{Lavrijsen+2020}& 2020& NOMAD, ImFil, SnobFit, and BOBYQA& VQE& Unit test&Finding the best optimizer for hybrid algorithms running on NISQ hardware&  \\ & & & & &&  \\ \hline
  \cite{Roth+2025}& 2025& Vibration sensor, Industrial Images, Price, Sensor Time Series dataset& Using sQUlearn library to support a variety of QML algorithms& ACC: 0.94, Ballanced ACC: 0.80, Mean Absolute Perecentage Error: 0.12-0.15, Mean Absolute Scaled Error (MASE): 0.53-0.54&AQML for industrial use cases& \\ \cline{1-6} \cite{Altares+2024}& 2024& MRI and X-ray Images& PCA+QSVM, CAE+ QSVM, PCA+ MLP& ACC: 0.684-0.967&Automatic generation and training of quantum-inspired classifiers&  \multirow{14}{*}{\raisebox{-10em}{\rotatebox{90}{\fontsize{10}{8}\selectfont \textbf{Category D: AQAI}}}}\\ \cite{Rybotycki+2024}& 2024& Simulated Dataset& AQMLator& Mean accuracy and Silhouette scores&Automatically propose and train the quantum layers of an ML model&  \\ \cline{1-6} \cite{Incudini+2024}& 2024& Proton collision dataset& Designing an algorithm for automated quantum kernel and meta-heuristic optimization& Cost function&Automatic  discovery of
quantum kernels&  \\ \cline{1-6}\cite{Sankari+2023}& 2023& 200 Neonatal Fundus Images& ResNet50, DarkNet19, CNN, SVM, Reduced Error Pruning
(REP) tree, K-Star, LogitBoost, QSVM& ACC: 72-95.5, Sensitivity: 72-93, Specificity: 70-100, PPV: 68-98, NPV: 71-93.33, AUC: 0.72-0.97&Automated detection of retinopathy of prematurity&  \\ \cline{1-6} \cite{Koike+2022}& 2022& Simulated Dataset& AutoAnsatz& ACC (Various Ansatz): 0.15-0.8, ACC (Various machine learning algorithms): 0.44-0.95&AQML for Wi-Fi Integrated Sensing and Communications&  \\ \cline{1-6} \cite{Gomez+2022}& 2022& Energy price dataset& qGANs& $\text{µ}_{\text{RE}}$: 0.2434-1.1412, $\text{$\sigma$}_{\text{RE}}$: 0.0269-0.6072, $\text{µ}_{\text{KS}}$: 0.1104-0.3420,  $\text{$\sigma$}_{\text{KS}}$: 0.0305-0.1676&Automated
circuit architecture search framework&  \\ \cline{1-6} \cite{Altares+2021Automatic}& 2021& Moons synthetic non-linear dataset and realistic benchmarks related to Parkinson, IoT irrigation, Drug classification& Genetic algorithm& ACC: 0.65-1.00&Automatic generation of optimal ad-hoc ans\"{a}tze&  \\ \hline
 \end{tabular}}
\label{tab:literature-classification}
\end{center}
\end{table*}
\section{Threats to Validity}

As with any systematic review, this study is subject to several potential threats to validity. In the following subsections, we review the threats to construct validity, internal validity, external validity, and conclusion validity.

\subsection{Construct Validity}

Construct validity refers to whether the study accurately captures the concepts it intends to investigate. In our case, this pertains to how effectively the selected search terms and classification schema represent the domains of QAI, QSE, AQSE, and AQAI. Although we curated the keyword list based on expert knowledge gained by reviewing the relevant literature and iteratively refined it, there is still a possibility that relevant papers using alternative terminology were unintentionally excluded. Furthermore, our four-category taxonomy is based on conceptual boundaries that may not be universally adopted in the literature.

\subsection{Internal Validity}

Internal validity concerns the processes used to identify and select studies. While we carefully defined inclusion and exclusion criteria and applied them consistently, the process may still be influenced by researcher bias in interpreting the relevance of borderline cases. To mitigate this, we followed a structured selection process and performed multiple iterations to verify consistency. Additionally, the inclusion of arXiv preprints, though justified by the fast-evolving nature of the field, may lead to differences in quality because they haven’t been peer-reviewed.

\subsection{External Validity}

External validity refers to the generalizability of our findings. Since our review focuses only on publications from 2020 to 2025, some earlier foundational studies may have been omitted. While these restrictions help in capturing the most recent state-of-the-art, they may limit the completeness and historical perspective of our review.

\subsection{Conclusion Validity}

Conclusion validity relates to the correctness and credibility of the conclusions drawn. While we applied systematic procedures during analysis and classification, the interpretation of content and assignment to categories inevitably involves human judgment. Although we attempted to be as objective as possible, some degree of subjectivity remains.

\section{Conclusion and Future Directions}\label{sec:conclusion-future-directions}
In this paper, we have presented a systematic review of research studies published in several identified venues from 2020 to 2025, related to quantum software and AI engineering, specifically the four categories of QSE, QAI, AQSE, and AQAI. The number of cited research work that are directly related to the stated categories is 48. Our work has highlighted the potential of QC to advance software engineering and AI by enhancing computational capabilities, while emphasizing the crucial role of automation in driving these developments.

Overall future directions for this research will focus on addressing gaps in the field, including the critical need for technologies that can automatically transform, optimize, and deploy code across various quantum processors with different hardware technologies, architectures, and models of computation. To tackle this challenge, we plan to leverage machine learning and MDSE in an automated manner, with the goal of generating code that supports different SDKs to realize machine learning-enabled systems that run on hybrid quantum and classical hardware resources.

\section{Acknowledgment}
This work is funded by a grant (Q-Dev) from the Colorado Office of Economic Development and International Trade (OEDIT). In addition, in preparing this work, we used generative AI tools (OpenAI's Chat-GPT) to enhance the content throughout the paper.

\bibliographystyle{ieeetr}
\bibliography{references}

\newpage
\begin{IEEEbiography}[{\includegraphics[width=1in,height=1.25in,clip,keepaspectratio]{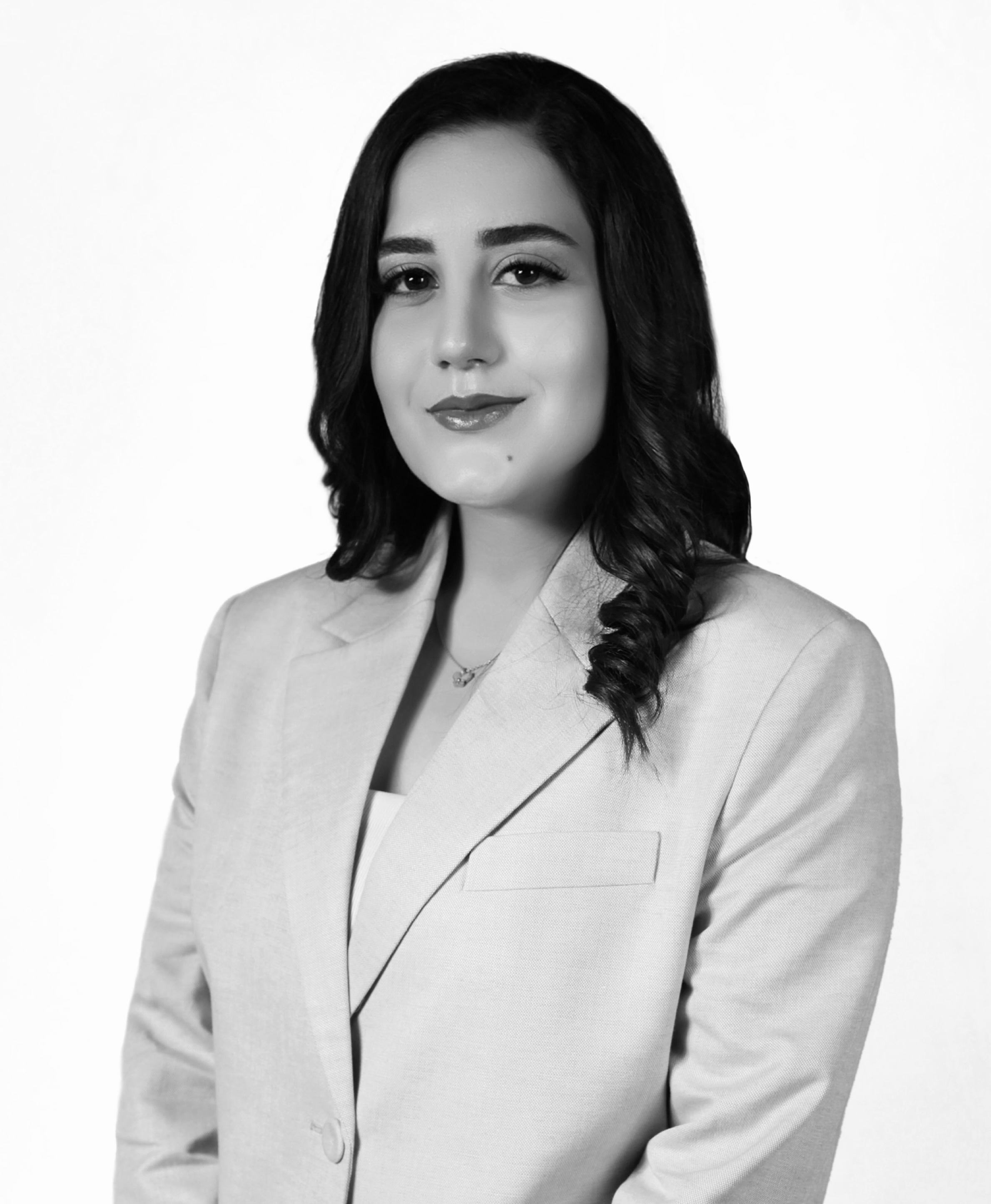}}]{Nazanin Siavash} Nazanin Siavash is a Ph.D. student and Graduate Research Assistant at the Quantum-Classical AI and Software Engineering (QAS) Lab in the Department of Computer Science at the University of Colorado Colorado Springs (UCCS). She received her B.Sc. in Electrical Engineering from Shahid Beheshti University, Tehran, Iran, in 2023. Her research focuses on model-driven engineering for hybrid quantum-classical artificial intelligence systems. During her undergraduate studies, she also served as a Research Assistant and contributed as a reviewer for the IEEE Transactions on Energy Conversion.
\end{IEEEbiography}
\begin{IEEEbiography}[{\includegraphics[width=1in,height=1.25in,clip,keepaspectratio]{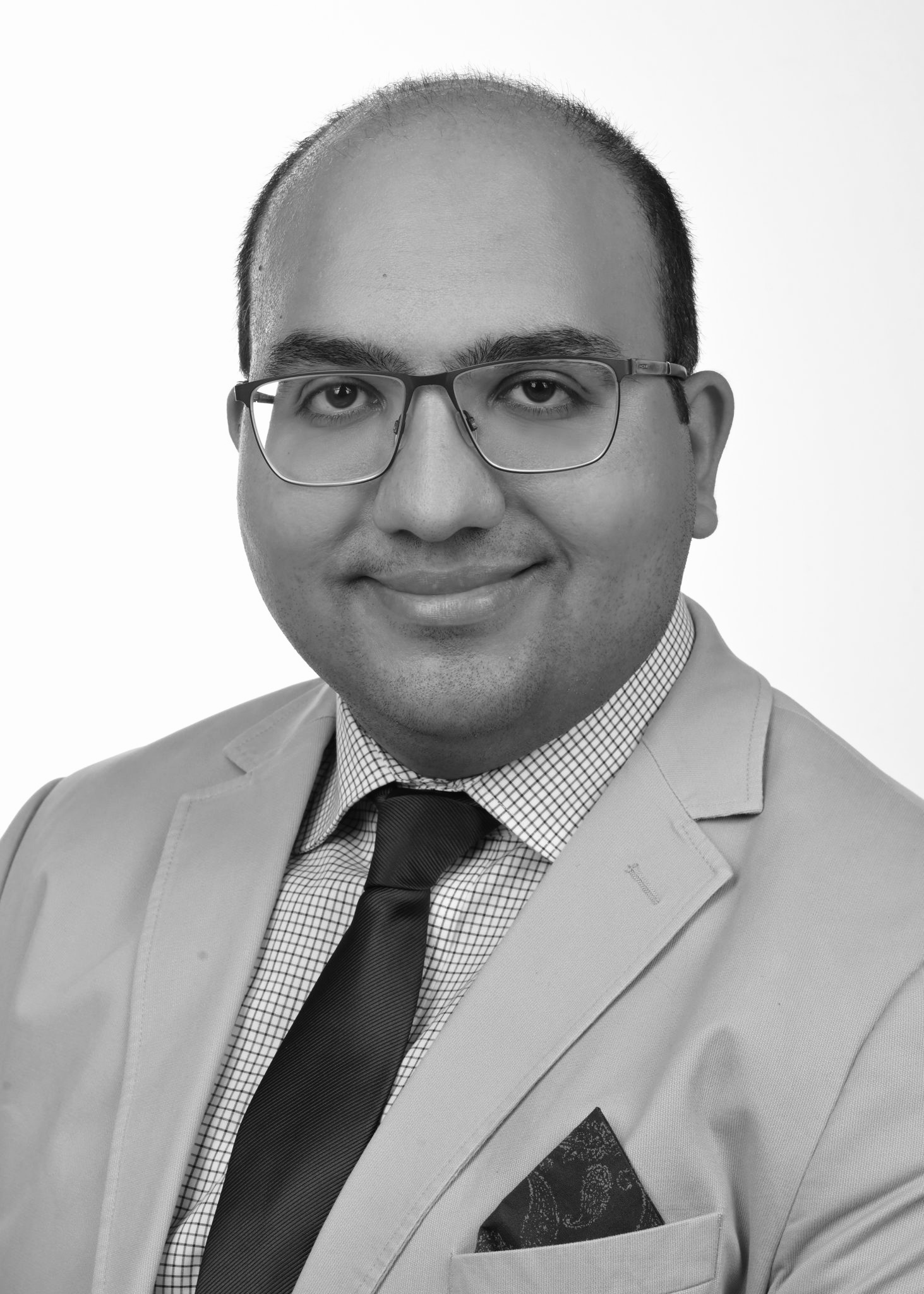}}]{Armin Moin} Dr. Armin Moin is a Tenure-Track Assistant Professor and Director of the Quantum-Classical AI and Software Engineering (QAS) Lab at the Computer Science (CS) Department of the University of Colorado Colorado Springs (UCCS). Before starting his faculty position, he worked as a Postdoctoral Scholar-Employee at the CS Department of the University of California, Santa Barbara (UCSB) in the U.S. and as a Postdoctoral Scholar at the CS Department of the University of Antwerp and FlandersMake in Belgium. Dr. Moin obtained his Ph.D. in CS from the Technical University of Munich (TUM), Germany, one of the world’s top universities, in 2022. He also has a Master’s in CS and an Executive MBA in Innovation and Business Creation. His research focuses on the intersection of Artificial Intelligence (AI) and Software Engineering (SE), particularly AI4SE and SE4AI, with an emphasis on hybrid quantum-classical computing. Grants from various sources, including the U.S. National Science Foundation (NSF) and the Colorado Office of Economic Development and International Trade (OEDIT), have supported his QAS Lab. Besides conducting research and teaching at UCCS, Dr. Moin reviews top academic conferences and journals.
\end{IEEEbiography}

\EOD

\end{document}